# Quantum electronic transport of topological surface states in $\beta$-Ag$_2$Se nanowire

*Jihwan Kim,[1,†] Ahreum Hwang,[1,†] Sang-Hoon Lee,[2] Seung-Hoon Jhi,[2] Sunghun Lee,[1] Yun Chang Park,[3] Si-in Kim,[1] Hong-Seok Kim,[4] Yong-Joo Doh,[4,\*] Jinhee Kim,[5,\*] Bongsoo Kim[1,\*]*

[1]Department of Chemistry, KAIST, Daejeon 34141, Korea

[2]Department of Physics, Pohang University of Science and Technology, Pohang 37673, Korea

[3]Department of Measurement and Analysis, National Nanofab Center, Daejeon 34141, Korea

[4]Department of Applied Physics, Korea University Sejong Campus, Sejong 30019, Korea

[5]Korea Research Institute of Standards and Science, Daejeon 34113, Korea

[†]These authors contributed equally to this work.

*E-mail: (B.K.) bongsoo@kaist.ac.kr; (J.K.) jinhee@kriss.re.kr; (Y.J.D) yjdoh@korea.ac.kr

**ABSTRACT**

Single-crystalline $\beta$-Ag$_2$Se nanostructures, a new class of 3D topological insulators (TIs), were synthesized using the chemical vapor transport method. The topological surface states were verified by measuring electronic transport properties including the weak antilocalization effect, Aharonov-Bohm oscillations, and Shubnikov-de Haas oscillations. First-principles band calculations revealed that the band inversion in $\beta$-Ag$_2$Se is caused by strong spin-orbit coupling and Ag-Se bonding hybridization. These extensive investigations provide new meaningful information about silver-chalcogenide TIs that have anisotropic Dirac cones, which could be useful for spintronics applications.

Topological insulators (TIs) are bulk insulators with metallic surface states that are topologically protected by time-reversal symmetry.[1] The strong spin-orbit coupling (SOC) inherent to TIs causes the spin orientation of the surface electrons to be locked perpendicular to their translational momentum, resulting in the suppression of electron backscattering from nonmagnetic perturbations.[2] The formation of spin-textured metallic edge states in TIs enables highly coherent charge and spin transport, making TIs promising for spintronics applications.[3] Combinations of TIs with conventional superconductors can also provide useful platforms for creating and manipulating emergent particles such as Majorana fermions,[4,5] which are essential for topological quantum computers.[6]

The topological surface states (TSSs) in three-dimensional (3D) TIs form two-dimensional (2D) electron gases populated by massless Dirac fermions with spin-momentum locking.[7] Photoemission spectroscopy[8-11] and electrical transport measurements[12-20] have been used to investigate the TSSs in 3D TIs such as $Bi_xSb_{1-x}$, $Bi_2Se_3$, and $Bi_2Te_3$, which have isotropic Dirac cones. TSSs with highly anisotropic Dirac cones,[21] which would be useful for realizing ideal quantum wires exhibiting one-directional spin and electron transport on their surfaces,[22] are theoretically expected in silver chalcogenides ($Ag_2Se$ and $Ag_2Te$).[23] Experimental works, however, have been reported only quite recently on the electronic transport of anisotropic Dirac fermions in silver chalcogenides TIs.[24,25]

Herein, we report the observation of TI properties in nanowires (NWs) and nanoribbons (NRs) of $\beta$-$Ag_2Se$. We synthesized single-crystalline NWs, NRs, and nanoplates (NPLs) of $\beta$-$Ag_2Se$ using the chemical vapor transport (CVT) method and measured quantum electrical transport properties of $\beta$-$Ag_2Se$ NWs and NRs at low temperatures. Negative magnetoconductance (MC) was observed, indicating the presence of the weak antilocalization (WAL) effect and implying the existence of strong SOC. When a magnetic field was applied parallel to the NW axis, the *β*-Ag$_2$Se NW device exhibited highly periodic MC oscillations, implying the existence of the surface electronic states. When a magnetic field was applied perpendicular to the NW axis, quantum magnetic oscillations, the so-called Shubnikov-de Haas (SdH) oscillations, were observed. The analysis of the SdH oscillations indicates the presence of a nonzero Berry phase due to the TSS of the *β*-Ag$_2$Se nanostructure. First-principles band structure calculations confirmed the existence of the TSS.

Because of their anisotropic space and violation of rotational invariance, anisotropic TIs can be potentially employed for observation of exotic fermion through its spaces.[21,26] Since the in-depth studies for anisotropic TI are quite rare, detailed investigations of these TIs could provide valuable insight into their exotic electrical transport properties, and may allow their possible applications for spintronics and quantum information applications.

## RESULTS AND DISCUSSION

Single crystalline *β*-Ag$_2$Se nanostructures were synthesized on a *c*-Al$_2$O$_3$ substrate using the single-step CVT method (Figure S1, Supporting Information), which enables high quality stoichiometry and assures clean surfaces of the nanostructures.[27] The cross-section of the NW had a hexagonal shape, with a diameter of 100 – 250 nm (Figure 1). NRs and NPLs were also synthesized on the same substrates. The widths of the NRs were about 0.4 – 3 μm and their lengths were 3 – 40 μm, while NPLs had edge lengths of 3 – 10 μm.

We performed detailed structural and compositional analyses of the *β*-Ag$_2$Se NWs using cross-sectional transmission electron microscopy (TEM). The high-resolution TEM (HRTEM) image and selected-area electron diffraction (SAED) pattern reveal the $[10\bar{1}]$ growth direction and the single crystalline nature of the NW (Figure 1e, f). Lattice spacings of 0.719 nm and 0.383 nm agree well to those of the (010) and (101) planes of an orthorhombic *β*-

Ag$_2$Se crystal structure, respectively. All peaks in the X-ray diffraction (XRD) pattern of the synthesized sample are indexed to the orthorhombic $\beta$-Ag$_2$Se phase (space group P2$_1$2$_1$2$_1$, JCPDS card No. 01-071-2410) with a set of lattice constants of a = 4.333 Å, b = 7.062 Å, c = 7.764 Å (Figure S2, Supporting Information). Additional TEM analyses and TEM energy dispersive X-ray spectroscopy (EDS) measurements also confirmed that all of the synthesized nanostructures are single-crystalline with a Ag/Se atomic ratio of 2:1 (Figure S3, Supporting Information). Details of the NW growth, device fabrications, and the electrical measurements are described in Methods.

We measured five representative $\beta$-Ag$_2$Se NW and NR samples to observe quantum electronic transports. Physical parameters of the samples (**D1** ~ **D5**) are listed in Table S1 (Supporting Information). Figure 2a shows a SEM image of Sample **D1**. Angle-dependent MC curves are obtained at $T$ = 2.0 K (Figure 2b). Magnetic field was applied at four different angles ($\theta$). The observed MC is negative, a typical feature of WAL,[28] and it becomes more negative as the perpendicular component of the magnetic field increases (with increasing $\theta$). This enhancement of negative MC indicates that the TSS is effective here instead of bulk.[16,29] Since significant negative MC already exists when $\theta$ = 0°, which is attributable to the strong SOC in the bulk portion, we extracted the surface state contribution by subtracting the axial MC component from the total MC, $\Delta G(\theta, B) = G(\theta, B) - G(\theta = 0°, B)$. When the $\Delta G$ curves were replotted as functions of the perpendicular magnetic field $B\sin\theta$, they mapped onto a single curve at low magnetic fields (Figure 2c). This extracted term is attributed to the surface states contribution.[16,17,29]

The low-field $\Delta G(\theta = 90°, B)$ curve was fitted to the 2D WAL model,[28] resulting in a phase coherence length $L_\varphi$ of 270 nm at $T$ = 2.0 K (Figure S4, Supporting Information). This estimated value of $L_\varphi$, however, is much larger than the width ($w$) of the NW, 102 nm, invalidating the 2D theoretical assumption of $w \gg L_\varphi$. We therefore used a diffusive 1D localization model for further analysis,[17,30,31] including the electron-electron and spin-orbit scatterings:

$$\Delta G = \frac{\sqrt{2}e^2}{\pi \hbar} \frac{L_N}{L} \left[ \frac{3}{2} \frac{\text{Ai}\left(\frac{2L_N^2}{L_1^2}\right)}{\text{Ai}'\left(\frac{2L_N^2}{L_1^2}\right)} - \frac{1}{2} \frac{\text{Ai}\left(\frac{2L_N^2}{L_2^2}\right)}{\text{Ai}'\left(\frac{2L_N^2}{L_2^2}\right)} \right] \quad (1)$$

where $e$ is the electric charge, $\hbar$ is the reduced Planck's constant, $L$ is the length of the NW channel, $L_N$ is the Nyquist scattering length,[32] Ai is the Airy function, and Ai′ is its derivative. Here, $L_1 = \left(\frac{1}{L_\varphi^2} + \frac{4}{3L_{so}^2} + \frac{1}{3}\left(\frac{ewB}{\hbar}\right)^2\right)^{-\frac{1}{2}}$ and $L_2 = \left(\frac{1}{L_\varphi^2} + \frac{1}{3}\left(\frac{ewB}{\hbar}\right)^2\right)^{-\frac{1}{2}}$, where $L_{so}$ is the spin-orbit scattering length.[17] The fitting results are shown in Figure 2d, yielding $L_\varphi$ = 310 nm, $L_N$ = 240 nm, and $L_{so}$ = 200 nm for the $\Delta G$ curve obtained at $T$ = 2.0 K. $L_\varphi$ and $L_N$ are comparable to those obtained previously from a $Bi_2Se_3$ NR, while $L_{so}$ here is about five times larger than that of the $Bi_2Se_3$ NR.[17] A relatively large $L_{so}$ in the TSS of the $β$-$Ag_2Se$ NW indicates a more reduced SOC strength than in other TI NWs.[17, 29]

When the temperature increases, the $\Delta G$ curve broadens and diminishes due to increased thermal scattering (Figure 2d). The temperature dependences of the scattering lengths were obtained by fitting the $\Delta G(\theta = 90°, B)$ data to Eq. (1), exhibiting power-law behavior (Figure 2e). It is inferred that $L_\varphi$, $L_N$, and $L_{so}$ are proportional to $T^{-0.29}$, $T^{-0.31}$, and $T^{-0.27}$, respectively. Compared with the theoretical expectations,[30] according to which $L_\varphi$ and $L_N$ would decay with $T^{-1/3}$ in 1D and $T^{-1/2}$ in 2D, the exponents obtained in our experiment are significantly closer to the prediction by 1D localization theory than that by 2D localization theory.

Measurement of the axial MC by applying a magnetic field $B_{axial}$ along the NW axis confirmed the existence of metallic surface states in the $\beta$-Ag$_2$Se NW. After subtracting a smooth background signal from the MC curve (inset of Figure 3b), regular periodic oscillations with a differential conductance $\delta G$ were obtained (Figure 3a). The peak position of $\delta G$ matches well to a linear fit with the integer index $n$ (Figure 3b), indicating that the average period of the $\delta G$ oscillations is $\Delta B_{axial} = 0.40$ T. FFT analysis of the $\delta G(B_{axial})$ curve also resulted in the same oscillation period. Assuming that the NW has a rectangular cross section (139 nm width and 75 nm height, Figure 3c), the cross sectional area becomes $1.04 \times 10^{-14}$ m$^2$. Thus, the oscillation period corresponds to a magnetic flux $\Phi$ of $1.01\Phi_0$, where $\Phi_0 = h/e$ is the magnetic flux quantum. Similar $\Phi_0$-periodic oscillations were obtained using Sample **D5** (Figure S5, Supporting Information).

The periodic MC oscillations can be attributed to the phase-coherent propagation of the surface state electrons around the perimeter of the TI NW,[12,24,33] which is well known as the Aharonov-Bohm (AB) effect.[34] However, there is a significant difference[35,36]: although AB theory predicts MC minima occurring at integer multiples of $\Phi_0$, TI NW experiments have shown MC maxima. Moreover, the periodicity can decrease to half of $\Phi_0$, depending on the gate voltage.[13] A more plausible explanation than the AB effect is based on the formation of 1D subbands in the TI NW surface states and the nontrivial Berry phases of the surface-state electrons.[13,36] Since the 1D bandgap[36] has been estimated to be $\Delta_{1D} = hv_F/2(w + t) \approx 2.9$ meV, where $v_F$ is the Fermi velocity, $w$ is the width and $t$ is the height of NW,[23] the thermal broadening effect was neglected in our experiment (we assumed $\Delta_{1D} \gg k_B T$). In the regime of weak disorder and nonzero doping, theoretical calculations result in MC oscillations with a period of $\Phi_0$ and doping-dependent conductance maxima either at $n\Phi_0$ or at $(n + 1/2)\Phi_0$, where $n$ is an integer.[36] Since the circumference of the NW, $2(w + t)$, was much larger than the mean free path ($l_e \approx 20$ nm) and its conductance exhibited very weak gate dependence, our $\beta$-Ag$_2$Se NW was in the diffusive regime and away from the Dirac point, thus meeting the conditions assumed in the theoretical calculations.[35] The absence of $\Phi_0/2$-periodic oscillations in Figure 3d indicates that the TI NW was weakly disordered.

In addition to the above MC oscillations, magnetic quantum oscillations such as SdH oscillations constitute the most convincing evidence of surface electronic states.[37] Figure 4a shows magnetoresistance (MR) data obtained at a perpendicular magnetic field ($\theta = 90°$). After subtracting the smooth background, we obtained the SdH oscillation of the differential resistance $\Delta R$ as a function of $1/B$ (Figure 4b). The FFT analysis result (Figure 4b inset) indicates an oscillation frequency $B_f$ of 84.0 T. Since the oscillation period corresponds to $\Delta(1/B) = 1/B_f = 2\pi e/\hbar S_F$, where $S_F = \pi k_F^2$ is the cross-sectional Fermi surface area transverse to the applied field and $k_F$ is the Fermi wave vector, $k_F$ becomes 0.51 nm$^{-1}$. Thus, the 2D carrier concentration at the surface was $n_s = k_F^2/4\pi = 2.0 \times 10^{12}$ cm$^{-2}$ for the sample **D3**.

According to Lifshitz-Kosevich (LK) theory,[38] $\Delta R$ is given by $\Delta R = A\exp(-\pi/\mu B)\cos[2\pi(B_f/B + 1/2 + \gamma)]$, where $A$ is a temperature-dependent parameter, $\mu$ is the carrier mobility, and $2\pi\gamma$ is the Berry phase. The phase shift is expected to be $\gamma = 0$ for a conventional 2D electron gas and $\gamma = -0.5$ for 2D Dirac electrons.[39] Our experimental $\Delta R$ data are well fitted by the LK theory with $\mu = 560$ cm$^2 \cdot$V$^{-1} \cdot$s$^{-1}$ and $\gamma = -0.43$ (see the magenta curve in Figure 4b), supporting the existence of a Berry phase of $\pi$ due to the TSS. We obtain the elastic mean free path as $l_e = \hbar\mu k_F/e = 19$ nm and $k_F l_e = 9.4$. The SdH oscillations is also observed in the sample **D4** having a different geometry (Figure 4c). Sample **D4** also results in similar physical parameters of $n_s$, $\mu$, and $\gamma$ (see Table 1).

The phase factor γ was determined by analyzing the Landau levels (LLs), which are formed by the Landau quantization of the energy states in the magnetic fields. When the Δ$R$ minima in the SdH oscillations are taken to be integer indices ($N_{LL}$) of the LLs,[19] massless Dirac fermions are expected to satisfy $N_{LL} = B_f/B - 0.5$. Landau fan diagrams are shown in Figure 4d for two different samples. Linear fitting of the fan diagrams revealed a slope $B_f$ of 84.7 T and an intercept γ of −0.56 for the sample **D3.** Parameters for sample **D4** are $B_f$ of 122.0 T and γ of −0.42; these values are quite close to those estimated from the previous analyses based on the LK theory. Thus, our experimental results suggest that *β*-$Ag_2Se$ NWs contain massless Dirac fermions with Berry phases of π on their surfaces.

To clarify the nontrivial topological property of *β*-$Ag_2Se$, we performed the first-principles calculations based on the density functional theory.[40,41] The ionic character of Ag-Se bonding implies that the Ag-cation *s* states lie higher in energy than the Se-anion *p* states. The strong hybridization between Ag *d* states and Se *p* states, however, leads to the band inversion, as in $Ag_2Te$.[23] The strength of the *p-d* hybridization depends on the number of bonding neighbors of chalcogen atoms (seven for *β*-$Ag_2Se$ and eight for *α*-$Ag_2Te$). Thus the Ag *s* states are located at about −1.25 eV below the Fermi level, while the Se *p* states are above them (Figure 5b). Our band-structure calculation for a 15.53 nm-thick *β*-$Ag_2Se$ slab (Figure 5c) reveals that the bulk band gap is opened by the SOC and the surface states (red lines) with a linear dispersion develop inside the gap. Without SOC, the gap is closed and the surface states disappear. Figure 5d shows that the surface states are localized at the top and bottom surfaces of the *β*-$Ag_2Se$ slab. These calculation results are consistent with our experimental measurements, both providing evidences for the existence of TSSs in *β*-$Ag_2Se$ NWs.

The topological nature of *β*-$Ag_2Se$ can be confirmed by calculating the Wannier charge centers (WCCs) on the six faces of the first Brillouin zone. Figures 5e and 5f display the evolution of WCCs for the $k_y = 0$ and $k_y = \pi$ faces, respectively, by varying the pumping parameter $k_x$. As shown, the evolution curves (red solid curves) cross an arbitrary reference line (black dashed line) odd numbers of times on the $k_x$, $k_y$, and $k_z = 0$ faces and even numbers of times on the $k_x$, $k_y$, and $k_z = \pi$ faces. The odd numbers of crossings indicate that the Kramer pairs with time-reversal-invariant momenta exchange their partners during the pumping process, demonstrating the nontrivial topological nature of the material.[42,43] Our calculation results show that the topological index of $\beta$-Ag$_2$Se is 1;(0,0,0), supporting the suggestion of $\beta$-Ag$_2$Se as a new class of 3D TI.[44]

## CONCLUSIONS

We synthesized single-crystalline $\beta$-Ag$_2$Se nanostructures and studied their electrical transport properties. We observed the WAL effect, AB oscillations, and SdH oscillations, all of which support the existence of the conducting surface states and thus the TI nature of this material. Our analysis indicated that the $\beta$-Ag$_2$Se NWs had a nonzero Berry phase as a characteristic property of massless Dirac fermions. First-principles calculations confirmed the existence of TSSs in this material. As a new TI, $\beta$-Ag$_2$Se could provide a useful platform for studying the exotic electronic transport due to the anisotropic Dirac cone and for developing novel spintronic devices.

**EXPERIMENTAL METHODS**

**Synthesis.** Single crystalline $\beta$-Ag$_2$Se NWs, NRs and NPLs were synthesized in a horizontal hot-wall two-zone furnace with a 1 inch diameter inner quartz tube, as shown in Figure S1 (Supporting information). Ag$_2$Se powder of 0.02 g (Sigma-Aldrich) in an alumina boat was placed at the center of the upstream (US) zone. And sapphire (*c*-Al$_2$O$_3$) substrates were placed at ~ 30 cm downstream (DS) from the boat. The carrier Ar gas was supplied through a mass-flow controller at a rate of 30 sccm. The temperatures of the US zone and DS zone were maintained at 900 $^o$C and 950 $^o$C, respectively. The reaction time was 40 minutes, while the pressure was maintained at 3.6 Torr during the reaction.

**Characterization.** SEM images of $\beta$-Ag$_2$Se nanostructures were taken on a Nova 230 (FEI Co.). The XRD pattern of the as-grown nanostructures was recorded on a RIGAKU D/MAX-RC (12 kW) diffractometer operated at 30 kV and 60 mA with filtered CuK$_\alpha$ radiation. TEM, HRTEM images, SAED patterns and EDS spectra were taken by a JEOL JEM-2100F TEM (200 kV operation) and a FEI Tecnai G2 F30 (300 kV).

**Device fabrication and transport measurement.** A single $\beta$-Ag$_2$Se NW was transferred from a sapphire substrate to a SiO$_x$ coated Si substrate with pre-patterned Ti/Au pad. Conventional electron-beam lithography and RF sputtering were used to form Ti (15 nm)/Au (150 nm) electrodes. The e-beam resist, poly (methyl methacrylate) (PMMA 4% 950 K) was baked at 170 °C for 2 min. The MR was measured using a Physical Properties Measurement System (PPMS, Quantum Design Inc.) equipped with a rotating probe. We used a conventional four-probe measurement configuration combined with AC lock-in technique.

**Computational methods.** All calculations were performed using the first-principles methods based on the density functional theory as implemented in the Vienna *ab initio* simulation package (VASP)[45-47] (for $\beta$-Ag$_2$Se bulk) and the OpenMX codes[48-50] (for $\beta$-Ag$_2$Se slab). The exchange-correlation energy functional was treated with the generalized gradient approximation parameterized by Perdew, Burke and Enzelhof.[51] For bulk calculations, we used the projector augmented wave pseudopotentials[52] provided by VASP. We selected 400 eV as the energy cutoff for the plane-wave basis set. For slab calculations, we used the norm conserving pseudopotentials proposed by Morrison, Bylander, and Kleinman.[53] The pseudo-atomic basis orbitals are chosen as Ag7.0-s2p2d1 and Se7.0-s2p2d1. The lattice parameters of $\beta$-Ag$_2$Se were determined experimentally (a = 4.333 Å, b = 7.062 Å, and c = 7.764 Å)[54] and the first Brillouin zone was integrated using Γ-centered 16 x 10 x 9 k-point samplings for bulk and 9 x 6 x 1 for slab.

*ACKNOWLEDGMENTS*

B.K. acknowledges support of this work from National Research Foundation grant (NRF-2013R1A2A2A01069073). Y.J.D acknowledges support of this work from National Research Foundation of Korea through the Basic Science Research Program (Grant No. 2015R1A2A2A01006833). We are grateful to Kicheon Kang for useful discussions.

*Supporting information available*

**FIGURE CAPTIONS**

**Table 1.** Physical parameters of the surface states obtained from the LK analysis of SdH oscillations.

| Sample | $B_f$ (T) | $\mu$ (cm$^2$V$^{-1}$s$^{-1}$) | $\gamma$ | $n_s$ (cm$^{-2}$) | $k_F$ (nm$^{-1}$) | $l_e$ (nm) |
|---|---|---|---|---|---|---|
| **D3** | 84.0 | 560 | −0.43 | $2.0\times10^{12}$ | 0.51 | 19 |
| **D4** | 122 | 580 | −0.55 | $3.0\times10^{12}$ | 0.61 | 23 |

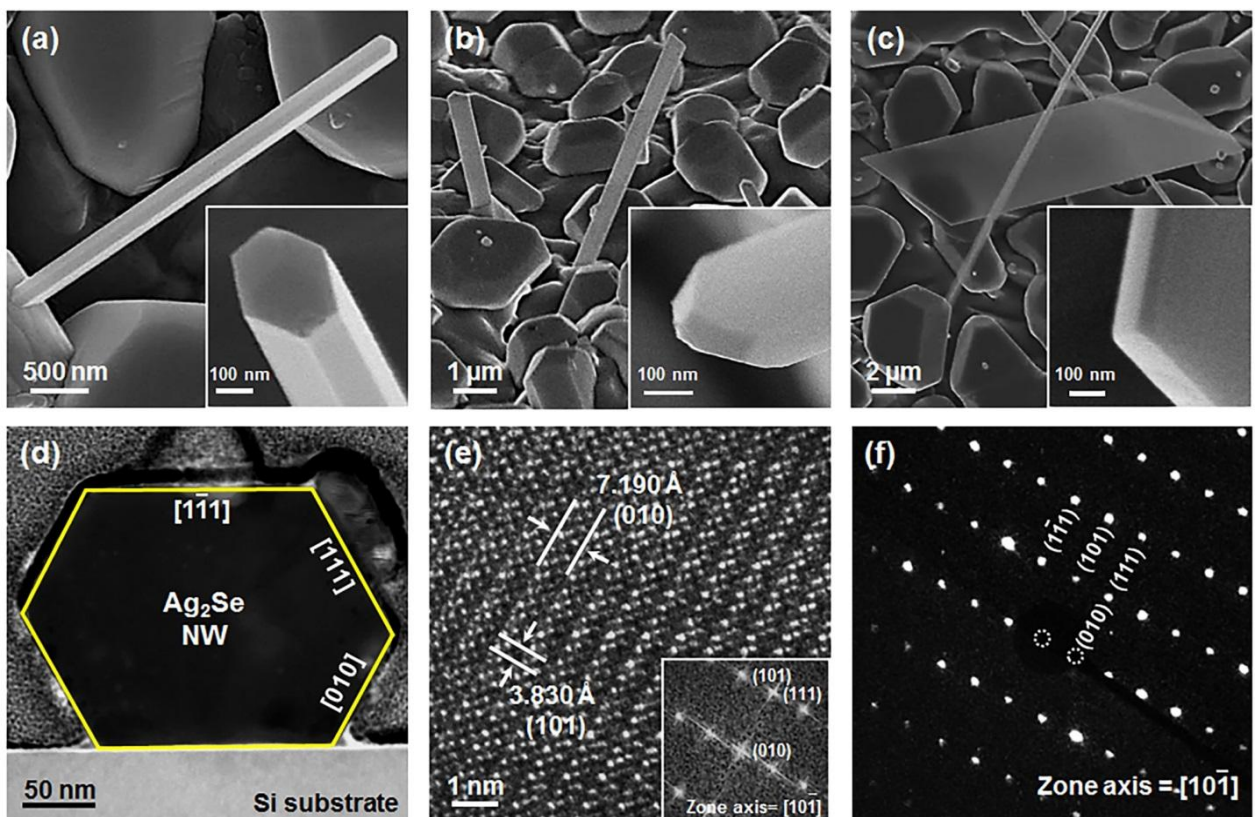

**Figure 1.** SEM images of as-synthesized $\beta$-Ag$_2$Se (a) NW, (b) NR and (c) NPL on a *c*-Al$_2$O$_3$ substrate. The insets in (a), (b) and (c) are magnified SEM images of the each nanostructures. (d) Cross-sectional TEM image of a $\beta$-Ag$_2$Se NW. (e) HRTEM image of (d). The FFT pattern (inset in (e)) is indexed to $\beta$-Ag$_2$Se with an orthorhombic lattice along the $[10\bar{1}]$ zone axis. (f) SAED pattern of a $\beta$-Ag$_2$Se NW along the $[10\bar{1}]$ zone axis.

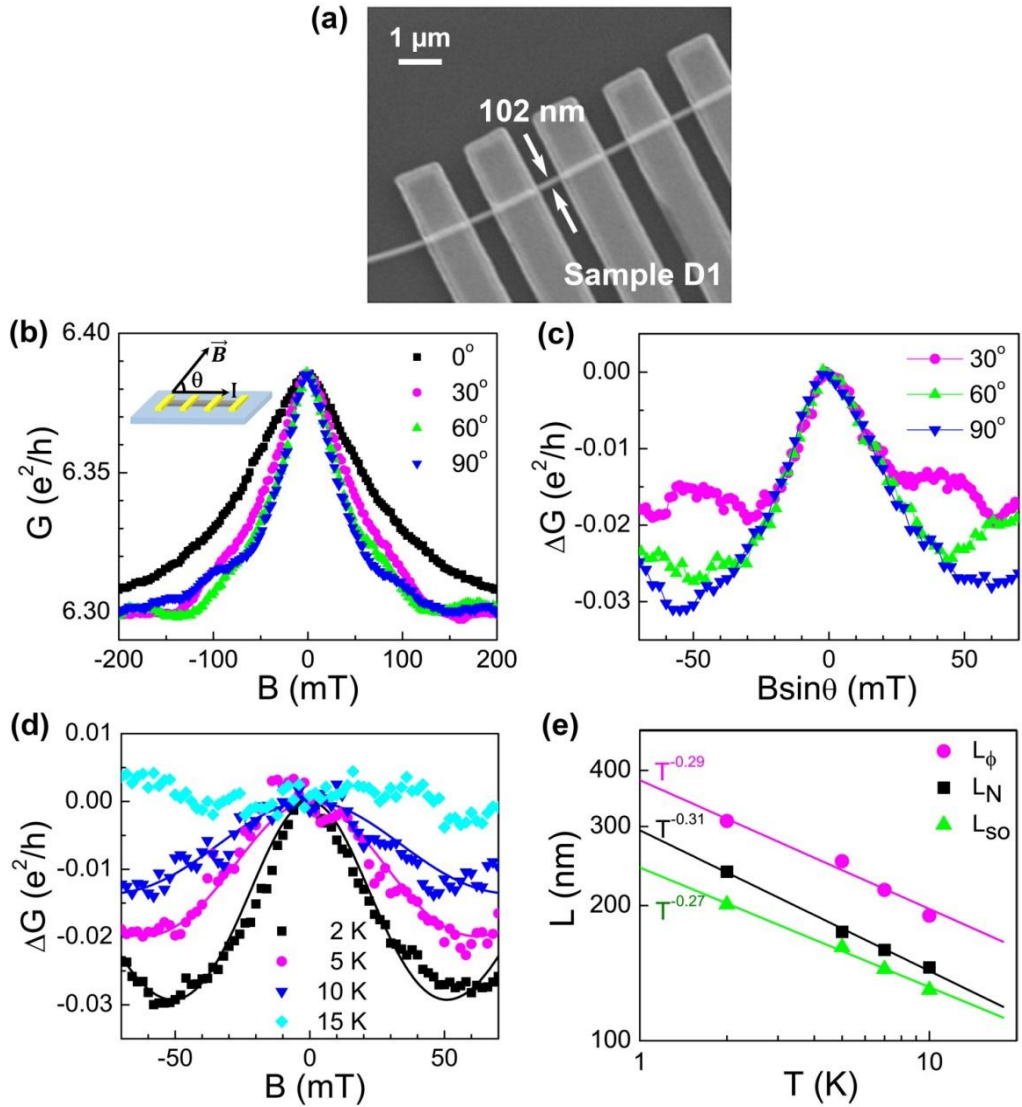

**Figure 2.** (a) SEM image of the $\beta$-Ag$_2$Se NW device (Sample **D1**) with a NW width of 102 nm (b) Angle-dependent MC curves at $T = 2.0$ K. Magnetic field was applied at various angles $\theta$ relative to the NW axis. Inset: Schematic of the measurement configuration. (c) Angle-dependent differential MC $\Delta G = G(\theta) - G(\theta=0^\circ)$, versus perpendicular component of magnetic field, $B\sin\theta$. (d) Temperature dependence of $\Delta G(B)$ curve (symbols). Solid curves are fitting results obtained by using 1D WAL theory. (e) Characteristic lengths of $L_\varphi$, $L_N$, and $L_{so}$ as functions of temperature. Solid lines are results of power-law fits.

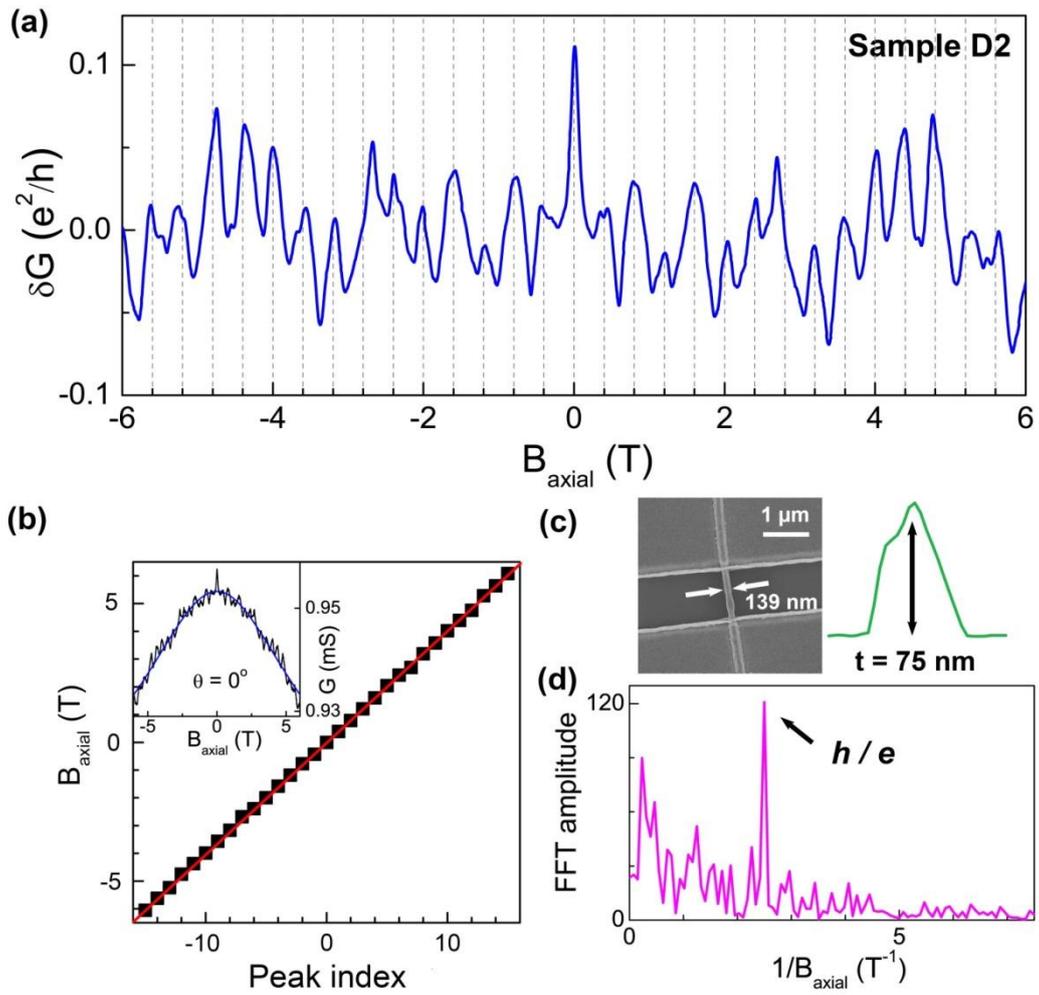

**Figure 3.** (a) Differential MC at $T$ = 2.0 K after subtracting smooth background. The dotted lines indicate an average period $\Delta B_{axial}$ = 0.40 T for the oscillation. (b) Magnetic field positions of $\delta G$ maxima versus integer index. Solid line is a linear fit. Inset: Raw data of axial MR obtained from sample **D2**. Blue curve is smoothly varying background. (c) Left : SEM image of a NW device with a NW width of 139 nm. Right: The height of the NW is 75 nm as determined by AFM. (d) FFT of $\delta G(B_{axial})$ curve in (a). Peak location, $1/B_{axial}$ = 2.499 T$^{-1}$, corresponds to $h/e$ oscillations.

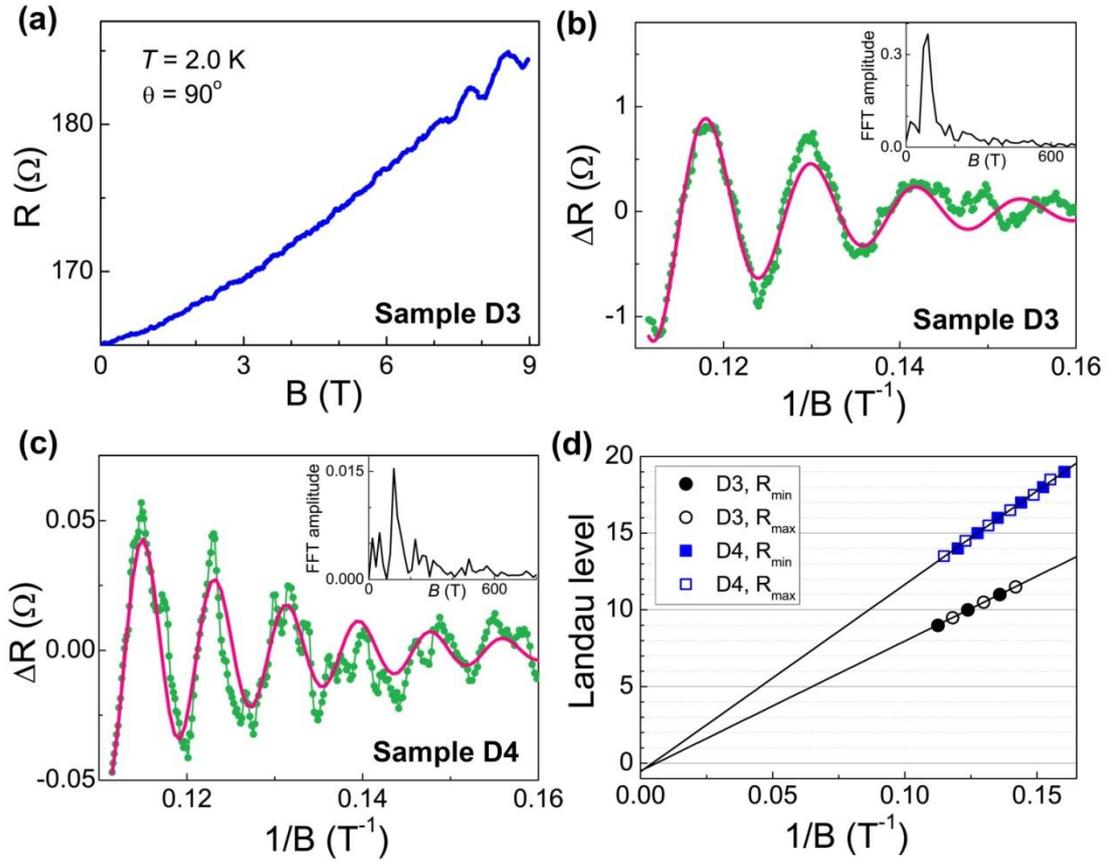

**Figure 4.** (a) MR data obtained from sample **D3** with a magnetic field perpendicular to the substrate. (b) Differential MR after subtracting smooth background. Solid magenta curve is a fit to LK theory (see text). Inset: FFT of $\Delta R$ with peak at $B_f = 84.0$ T. (c) $\Delta R$ obtained from sample **D4**. Solid magenta curve shows fitting result. Inset: FFT of $\Delta R$ with peak at $B_f = 122$ T. (d) Landau level index versus $1/B$ value obtained from sample **D3** (black) and **D4** (blue). $\Delta R$ minima (maxima) correspond to integer (half-integer) values of Landau level index. Lines are linear least square fits to data.

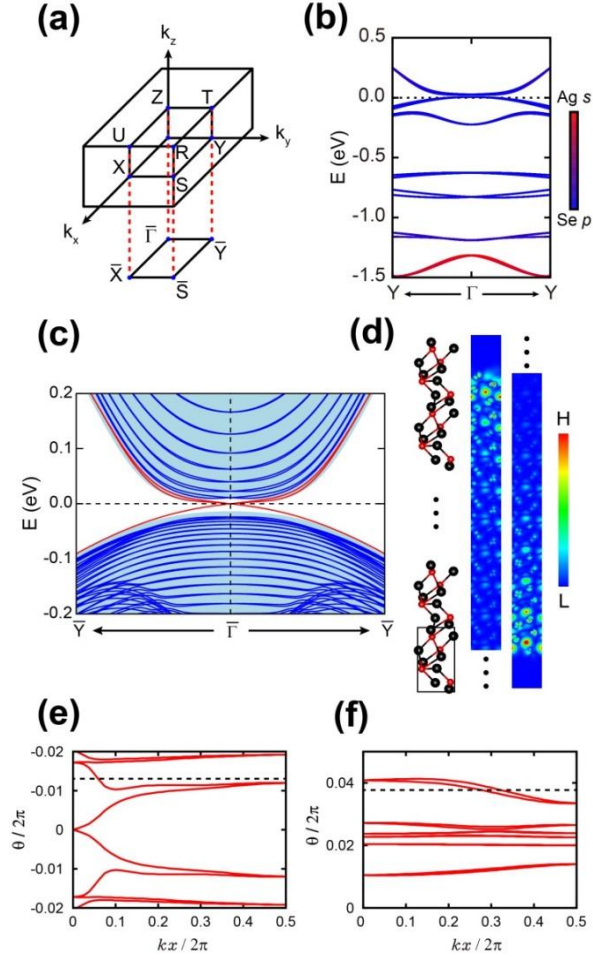

**Figure 5.** (a) First Brillouin zone of orthorhombic *β*-Ag$_2$Se and its 2D-projected surface Brillouin zone. (b) Contribution of Ag *s* (red) and Se *p* (blue) states to band structure of *β*-Ag$_2$Se. Bands from Ag *s* states (located about −1.25 eV) exhibit band-inversion feature. (c) Surface states in 15.53 nm-thick, *c*-axis normal *β*-Ag$_2$Se slab (red curve) with the (projected) bulk bands in blue curves (shaded light blue). (d) Part of *β*-Ag$_2$Se slab (left panel: Ag, black; Se, red) and surface-localized states of top (center panel) and bottom (right panel) surfaces at Γ. Evolution of Wannier charge centers in (e) $k_y = 0$ and (f) $k_y = \pi$ faces.

# Supporting Information

1. **Parameters of nanowires and nanoribbon**

2. **Experimental setup**

3. **XRD spectrum**

4. **TEM analyses of $\beta$-Ag$_2$Se nanowires and nanoplate**

5. **2D weak antilocalization fitting result**

6. **Aharonov-Bohm oscillation**

*Address correspondence to bongsoo@kaist.ac.kr, jinhee@kriss.re.kr, yjdoh@korea.ac.kr

Fax: +82-42-350-2810

**Table S1.** Parameters of β-Ag$_2$Se nanowires and nanoribbon samples.

| No. | Resistivity at 2 K (mΩ·cm) | Width (nm) | Height (nm) | Channel length (μm) |
| --- | --- | --- | --- | --- |
| **D1** | 5.2 | 102 | 95 | 0.75 |
| **D2** | 1.3 | 139 | 75 | 0.79 |
| **D3** | 0.15 | 227 | 150 | 3.69 |
| **D4** | 0.18 | 2770 | 200 | 11.8 |
| **D5** | 0.40 | 140 | 150 | 0.55 |

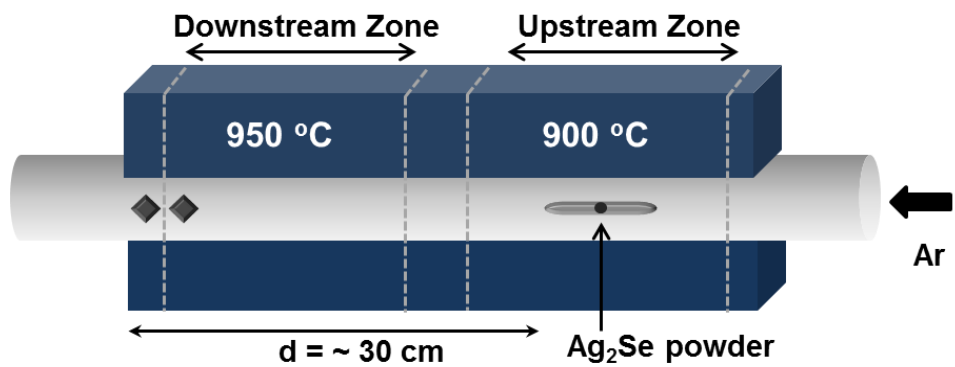

**Figure S1.** Experimental setup for the synthesis of *β*-Ag$_2$Se nanowires (NWs), nanoribbons (NRs) and nanoplates (NPLs).

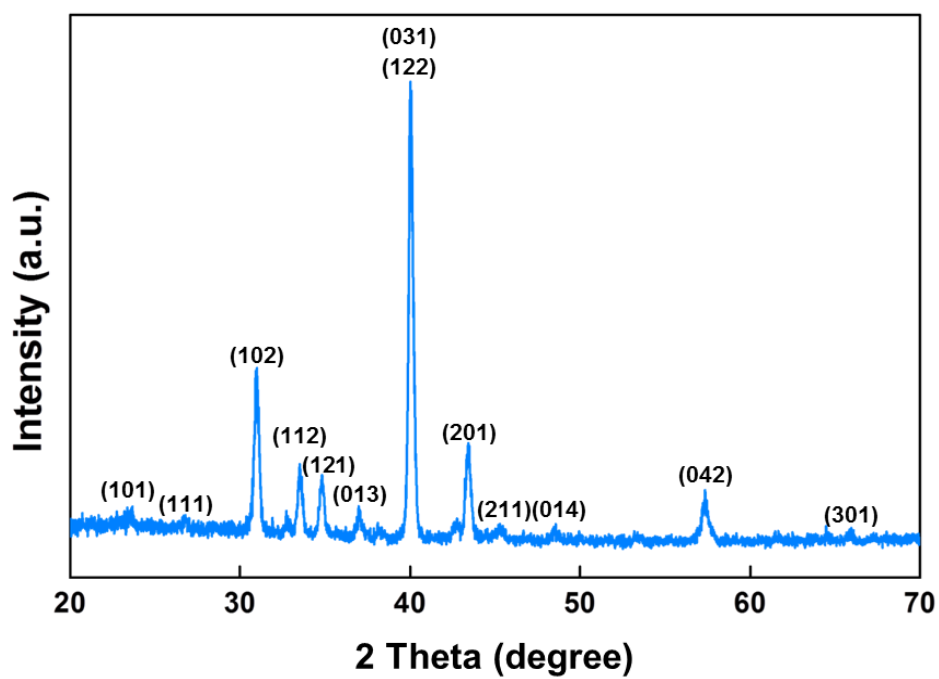

**Figure S2.** The X-ray diffraction (XRD) pattern obtained from as-grown NWs, NRs and NPLs on *c*-Al$_2$O$_3$ substrate. All the diffraction peaks are indexed to an orthorhombic *β*-Ag$_2$Se crystal structure (JCPDS card No. 01-071-2410).

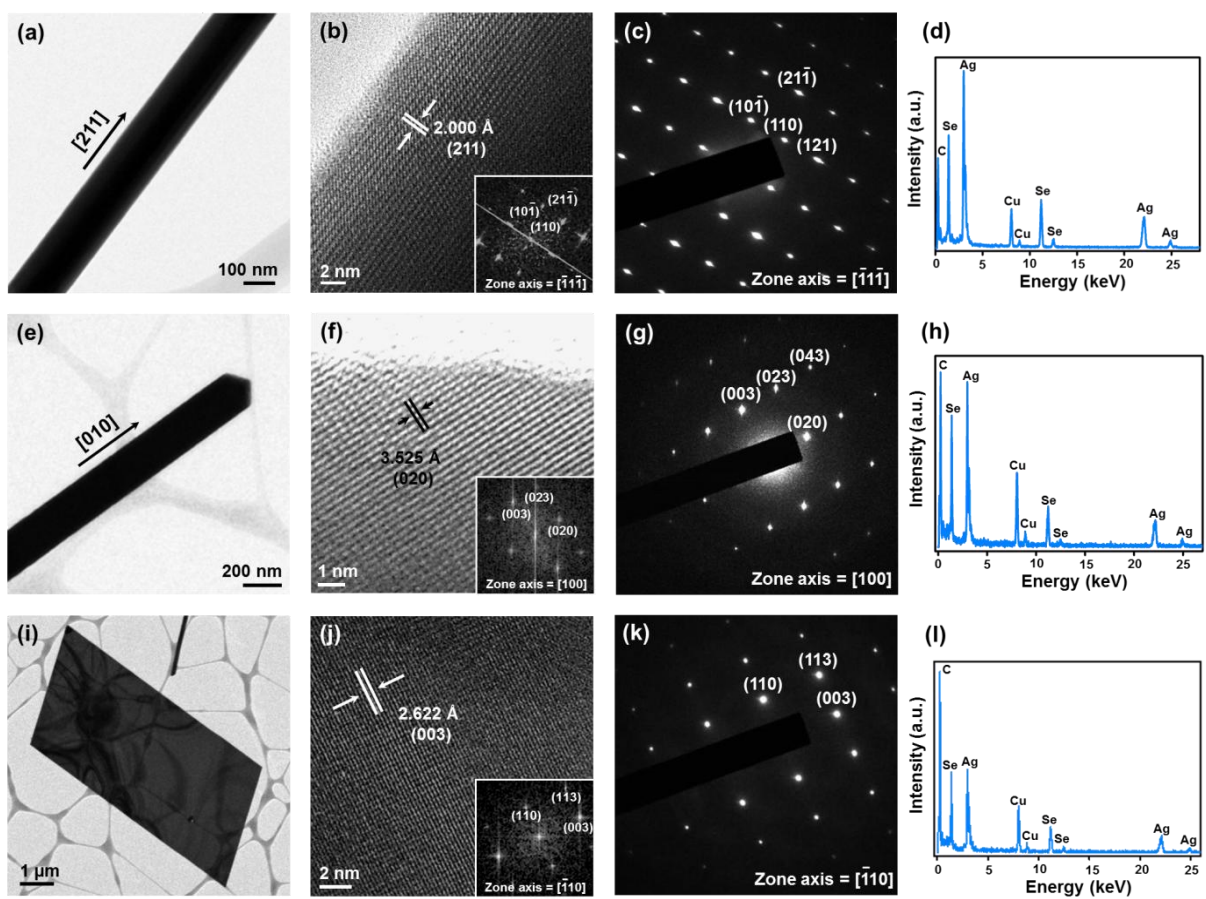

**Figure S3.** TEM images of *β*-Ag$_2$Se NWs and NPL.

First column ((a), (e), (i)): Low-resolution TEM images of the *β*-Ag$_2$Se NWs and NPL. Second column ((b), (f), (j)): High-resolution TEM images and FFT patterns (inset image) of *β*-Ag$_2$Se NWs and NPL. Lattice spacings of 0.200 nm, 0.353 nm, and 0.262 nm agree well to those of (211), (020), and (003) planes of an orthorhombic *β*-Ag$_2$Se crystal structure, respectively. Third column ((c), (g), (k)): The SAED patterns along the various zone axes. The patterns show the single crystalline nature of *β*-Ag$_2$Se nanostructures. Fourth column ((d), (h), (l)): TEM-EDS spectra of *β*-Ag$_2$Se NWs and NPL. The analyses of these results confirm that the NWs and NPL contain Ag and Se, in a ratio of 2:1. (Cu and C peaks are due to a

TEM grid)

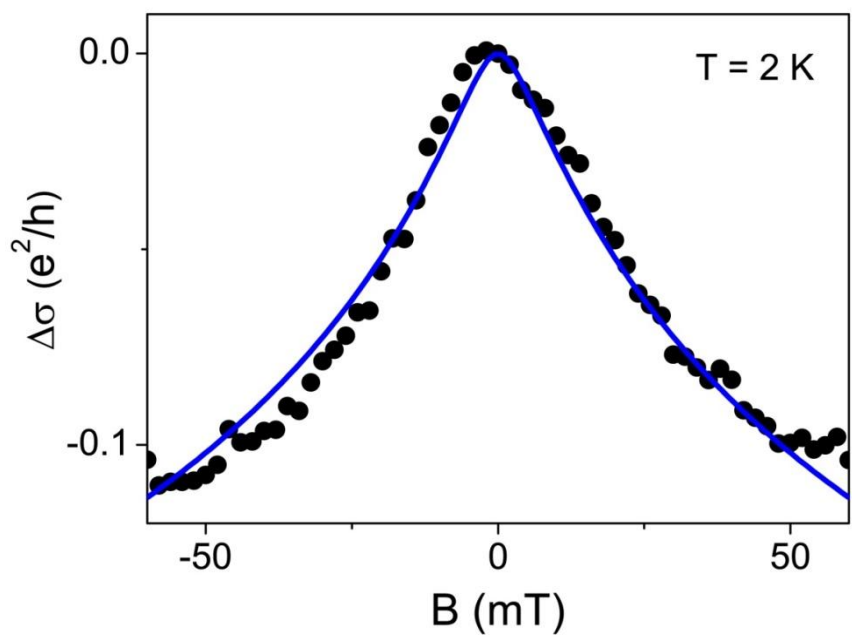

**Figure S4.** 2D differential magnetoconductance (MC) data (symbol) obtained from sample **D1** at $T$ = 2.0 K. Solid line is a fit to the 2D weak antilocalization (WAL) model.[1] The model expects 2D MC $\Delta\sigma = -(\alpha e^2)/(2\pi^2\hbar)[\ln(B_0/B) - \psi(1/2 + B_0/B)]$, where $\alpha$ is a prefactor, $B_0 = \hbar/(4eL_\varphi^2)$, $L_\varphi$ is a phase coherence length, and $\psi$ is a digamma function. A least square fitting results in $\alpha$ = 0.25 and $L_\varphi$ = 270 nm.

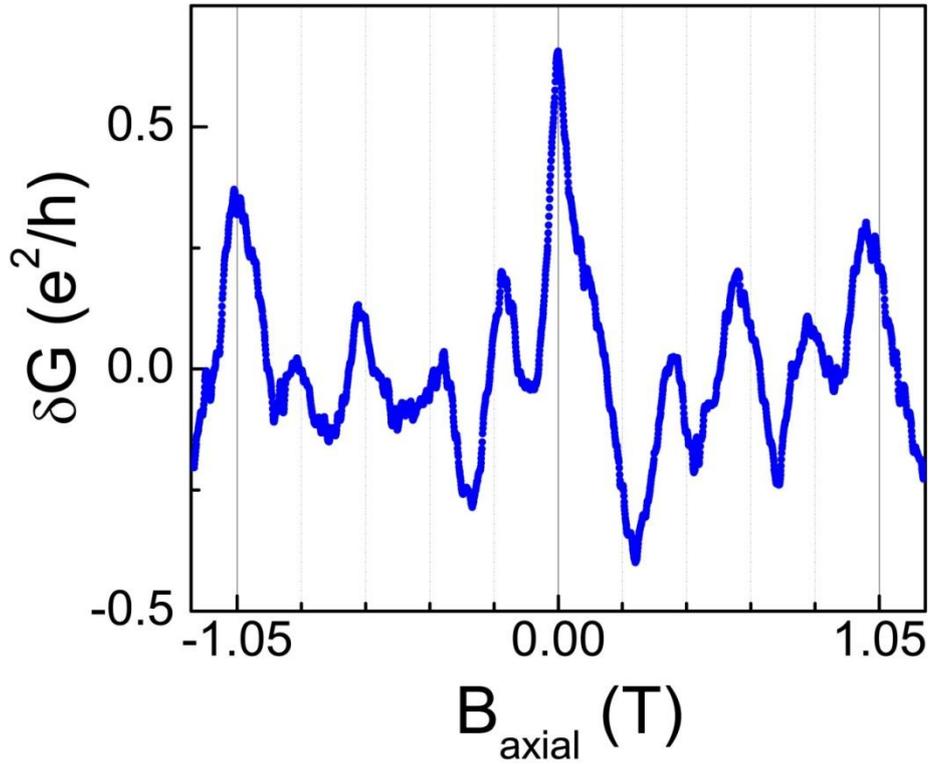

**Figure S5.** Differential MC obtained from sample **D5** at $T$ = 2.0 K. A smooth background MC was subtracted out. The dotted lines indicate an average period $\Delta B_{axial}$ = 0.21 T for the $\delta G$ oscillations. Assuming that the NW has a rectangular cross section, the area becomes 2.10 × $10^{-14}$ m$^2$. Thus, the oscillation period corresponds to a magnetic flux $\Phi$ = 1.07$\Phi_0$, where $\Phi_0$ = $h/e$ is the magnetic flux quantum. The flux can be overestimated by the approximation of the cross sectional shape.